\documentclass[aps,prl,preprintnumbers,twocolumn,superscriptaddress,nofootinbib]{revtex4-1}
\usepackage{amsmath,amsthm,amssymb,color,psfrag,url,latexsym,graphicx,epstopdf,slashed,xspace,hyperref,enumitem}
\usepackage{lineno}

\definecolor{darkred}{rgb}{0.6,0.0,0.0}
\definecolor{darkblue}{rgb}{0.0,0.0,0.5}
\definecolor{darkgreen}{rgb}{0.0,0.5,0.0}
\definecolor{brown}{rgb}{0.0,0.0,0.0}

\newcommand{\be}{\begin{equation}}
\newcommand{\ee}{\end{equation}}
\newcommand{\bea}{\begin{eqnarray}}
\newcommand{\eea}{\end{eqnarray}}

\begin{document}
\preprint{YITP-SB-19-30}
\title{Deep Learning Jet Substructure from Two-Particle Correlation}

\author{Kai-Feng Chen}
\email{kfjack@phys.ntu.edu.tw}
\affiliation{Department of Physics, National Taiwan University, Taipei, Taiwan}

\author{Yang-Ting Chien}
\email{yang-ting.chien@stonybrook.edu}
\affiliation{C.N. Yang Institute for Theoretical Physics, Stony Brook University, Stony Brook, NY 11794, USA}
\affiliation{Center for Frontiers in Nuclear Science, Stony Brook University, Stony Brook, NY 11794, USA}

\date{\today}

\begin{abstract}

Deciphering the complex information contained in jets produced in collider events requires a physical organization of the jet data. 
We introduce two-particle correlations (2PCs) by pairing individual particles as the initial jet representation from which a probabilistic model can be built. Particle momenta, as well as particle types and vertex information are included in the correlation. A novel, two-particle correlation neural network (2PCNN) architecture is constructed by combining neural network based filters on 2PCs and a deep neural network for capturing jet kinematic information. The 2PCNN is applied to boosted boson and heavy flavor tagging, and it achieves excellent performance by comparing to models based on telescoping deconstruction. Major correlation pairs exploited in the trained models are also identified, which shed light on the physical significance of certain jet substructure. 

\end{abstract}
\maketitle

In high energy collider events, hundreds or even thousands of particles are produced, and the understanding of their high-dimensional probability distributions can be a formidable task. An emergent structure consisting of collimated particles, referred to as jets, are typically observed. The kinematic distribution and many aspects of the internal structure of jets have been the testing ground of quantum chromodynamics (QCD) in perturbative calculations and non-perturbative modeling, with remarkable success witnessed in reasonably accurate descriptions of collider data via Monte Carlo (MC) simulations and analytic calculations.  However, the dynamical hadronization process which turns partonic degrees of freedom to hadronic degrees of freedom has not been fully understood and remains the holy grail of QCD.

In this letter two-particle correlations (2PCs) are explored as a representation of jet information, and such organization can illuminate the physics underlying jet formation from parton evolution to hadronization. This is one step beyond processing individual particle information by constructing particle pairs as basic information elements, from which probabilistic models can be built and physical analysis can be performed. The model includes not only the particle momenta for energy flow information, but also the electric charges and vertex information which are sensitive to hadronization as well as bottom quark decays. 

Note that, the number of 2PCs scales quadratically with the number of particles therefore it creates a redundancy in the jet representation. Moreover, an advantage can be gained from the effectiveness of how the relevant information is contained in each 2PC pair, and how these 2PC pairs build up significant features which one can identify and define concrete observables to probe. 

Modern computation power has made possible the rise of machine learning techniques, and many methods have been applied successfully on classification and regression problems in particle and nuclear physics, such as jet classification~\cite{
deOliveira:2015xxd,Baldi:2016fql,Kasieczka:2017nvn,Macaluso:2018tck,Butter:2017cot,
Louppe:2017ipp,Cheng:2017rdo,Egan:2017ojy,Chien:2018dfn,Fraser:2018ieu,Lin:2018cin,Andreassen:2018apy,
Andreassen:2019txo,Kasieczka:2019dbj,Pang:2016vdc,Pang:2019aqb,Lim:2018toa,Chakraborty:2019imr,Chen:2019uar}, 
correlation of particles~\cite{Komiske:2018cqr,Qu:2019gqs}, 
anomaly detection~\cite{Collins:2018epr,Farina:2018fyg,Blance:2019ibf,Roy:2019jae}, 
event generation~\cite{Paganini:2017dwg,deOliveira:2017pjk,Paganini:2017hrr},
and other tasks~\cite{Komiske:2017ubm,Andreassen:2019cjw}.
We will tackle classic classification problems such as boosted boson and heavy flavor jet tagging, as a way to discover and highlight certain jet properties which are relevant in these tasks. Specifically, the discrimination of two-prong jets ($W$ jets and Higgs jets from the $H\rightarrow b\bar b$ decay channel) and three-prong jets (fully hadronic top jets) against light quark $q$ ($q=u,d,c,s$ quark) jets, as well as $W^+$ versus $W^-$ \cite{Chen:2019uar}, and quark versus gluon jet discrimination, are studied. Excellent performance of 2PC-based neural network will be presented in all of the tasks. In particular, the network optimized for $W$ tagging successfully identifies the two-prong structure and isolation of $W$ jets by weighing strongly on these two features. The model behavior will be cross-checked by examining collinear and soft contributions from soft-drop \cite{Dasgupta:2013ihk,Larkoski:2014wba} and collinear-drop \cite{Chien:2019osu} constituents and their correlations. A combination of machine learning and physics analysis methods benefits significantly from the use of a physically organized and unbiased jet representation so that one can extract the physics features the model identifies. 

The analysis is performed with samples generated from MC simulations using MadGraph \cite{Alwall:2014hca} for hard scattering processes and \textsc{Pythia8} \cite{Sjostrand:2007gs} for parton shower and hadronization. Jets are defined using the anti-$k_T$ algorithm \cite{Cacciari:2008gp} implemented in \textsc{FastJet} 3~\cite{Cacciari:2011ma}, with $R=0.8$ for the studies of tagging high $p_T$ two or three-prong jets, and with $R=0.4$ for the studies of quark gluon discrimination. The high $p_T$ $R=0.8$ jets are generated using decays of hypothetical heavy $Z'$ bosons ($Z'\rightarrow W^+W^-, ZH, t \bar t, q \bar q$) with invariant mass fixed at 2 TeV, while the jets used in quark gluon discrimination are generated with the standard model QCD processes. For the samples generated using $Z'$ decays, jets are produced and reconstructed in the same kinematic region therefore the classification is not affected by the hard process kinematics. The truth particle information is passed through a Delphes~\cite{deFavereau:2013fsa} fast detector simulation and converted into particle flow candidates, with track, electromagnetic calorimeter and hadronic calorimeter information. 
A parametric model based on CMS detector~\cite{Chatrchyan:2008aa} at the Large Hadron Collider is introduced in the simulation.

\begin{figure}[t]
\begin{center}
\includegraphics[width=0.48\textwidth]{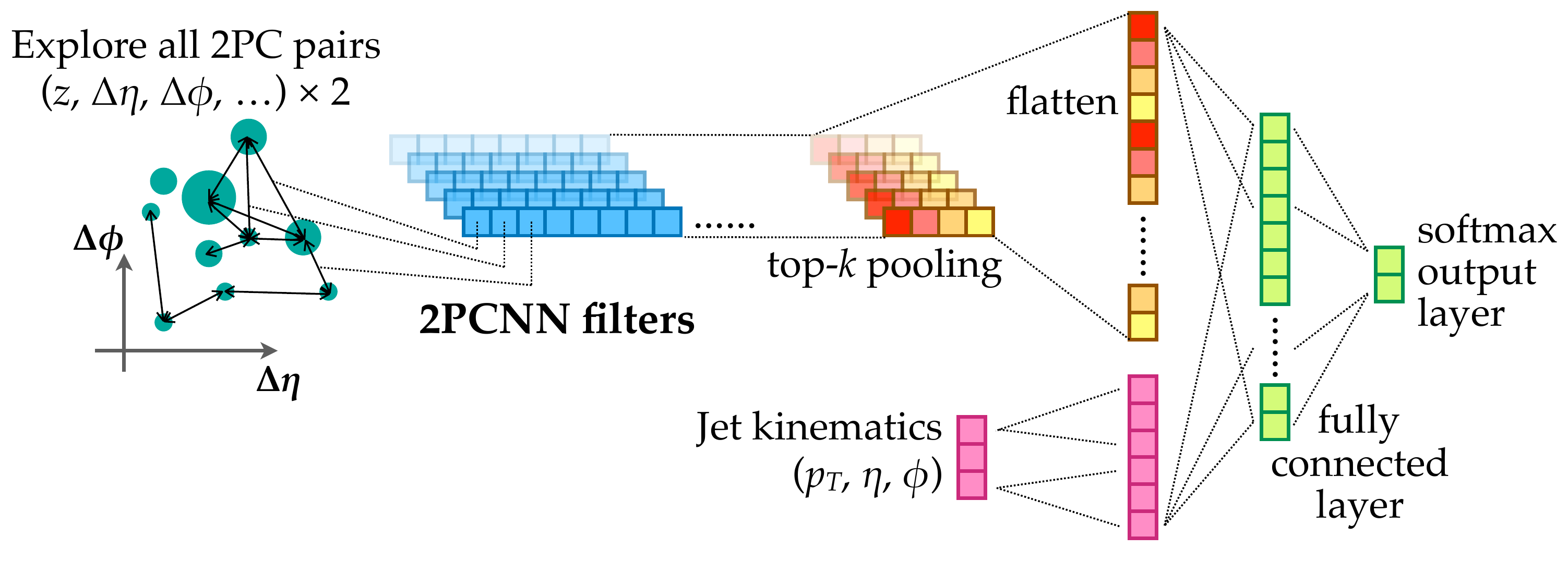}
\caption{The schematic view of the 2PCNN model. It processes two-particle correlations as inputs and uses filters with shared weights to benchmark the importance of each 2PC pair. The top-$k$-ranked filter outputs, together with jet kinematic information, are feed into a fully connected network for decision making. }
\label{fig:2PCNN_model}
\end{center}
\end{figure}

Two different sets of 2PC inputs are included. A basic set contains only the energy flow information\footnote{The energy flow input here includes infrared and collinear unsafe information.}, including the transverse momentum fraction $z = p_T^i/p_T({\rm jet})$, relative pseudorapidity $\Delta\eta = \eta^i - \eta({\rm jet})$, and relative azimuthal angle 
$\Delta\phi = \phi^i - \phi({\rm jet})$ of the jet constituents labelled by the index $i$.  Here $p_T^i$, $\eta^i$ and $\phi^i$ are the transverse momentum, pseudorapidity and azimuthal angle of particle $i$, respectively, and $p_T({\rm jet})$, $\eta({\rm jet})$ and $\phi({\rm jet})$ are the corresponding quantities of the jet. A rotation in $\Delta\eta$-$\Delta\phi$ coordinate system is performed to align the principle axis of the jet constituents horizontally. The other set of inputs contains the 2PCs of charged tracks, while the vertex position and the charge of each particle are introduced in addition. 

Based on the 2PC inputs, we design a two-particle correlation neural network (2PCNN)\footnote{The prototype 2PCNN example code and test samples are available from https://github.com/kfjack/2PCNN.} to model the probability distribution of jet particles (see FIG.~\ref{fig:2PCNN_model}), which is implemented using Keras \cite{chollet2015keras} with TensorFlow backend \cite{tensorflow2015-whitepaper}. Since the number of jet particles can vary, the 2PCNN layer is designed to handle inputs with variable sizes. Inspired by one of the key ideas from the convolutional neural network, the 2PCNN model implements a collection of filters\footnote{The filter consists of a fully-connected dense network with 2PCs as the input, processed with a hidden layer, and then a layer of single nodes as the output. We use the ReLU~\cite{nair2010rectified} activation function at each layer therefore the output can only be non-negative floating-point numbers.} with shared weights to process the input 2PC data. In the prototype model the number of filters is set to 64 to extract features from the energy flow information. The vertex and charge information is processed with a parallel 2PCNN layer containing 32 filters. Each filter processes and gives outputs to all input 2PCs. The filter outputs are then ranked according to their numerical values, and only the top-$k$ ranked 2PCs of each filter are kept as the inputs for the subsequent decision-making, fully connected network. In order to balance between performance and complexity, $k=4$ has been set; therefore the total number of output nodes is $256=64\times 4$, which is equal to the number of filters times~$k$.

Besides the 2PCNN layers, we use a dense network to include the jet kinematic information $p_T({\rm jet})$, $\eta({\rm jet})$ and $\phi({\rm jet})$ which is the baseline input for standard analysis. The outputs of the dense network and the 2PCNN layer are sent to another fully connected layer of 128 nodes (or 256 nodes if two 2PCNN layers are used), followed by two output nodes with softmax activation function for final decision. The model is optimized by minimizing a categorical cross-entropy loss function 
with the Adam optimizer~\cite{adam}. Input samples for each task are split into three subsets: one set consisting of 80k jets is used to optimize the weights in the model, and another set of 40k jets is used to validate if the model reaches its optimal performance. The other set of 40k jets is used for an independent measure of the model performance.

\begin{figure}[t]
\begin{center}
\includegraphics[width=0.23\textwidth]{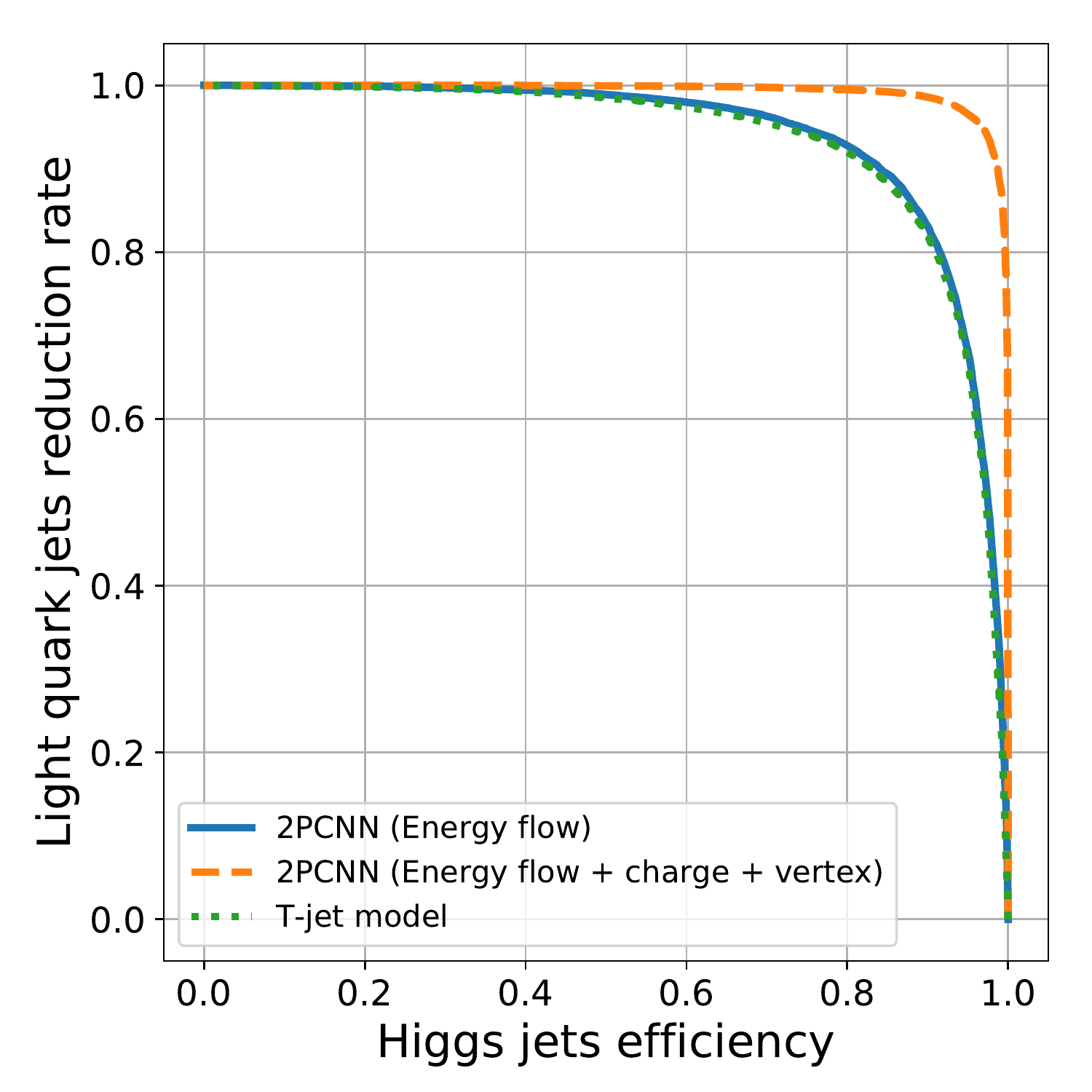}
\includegraphics[width=0.23\textwidth]{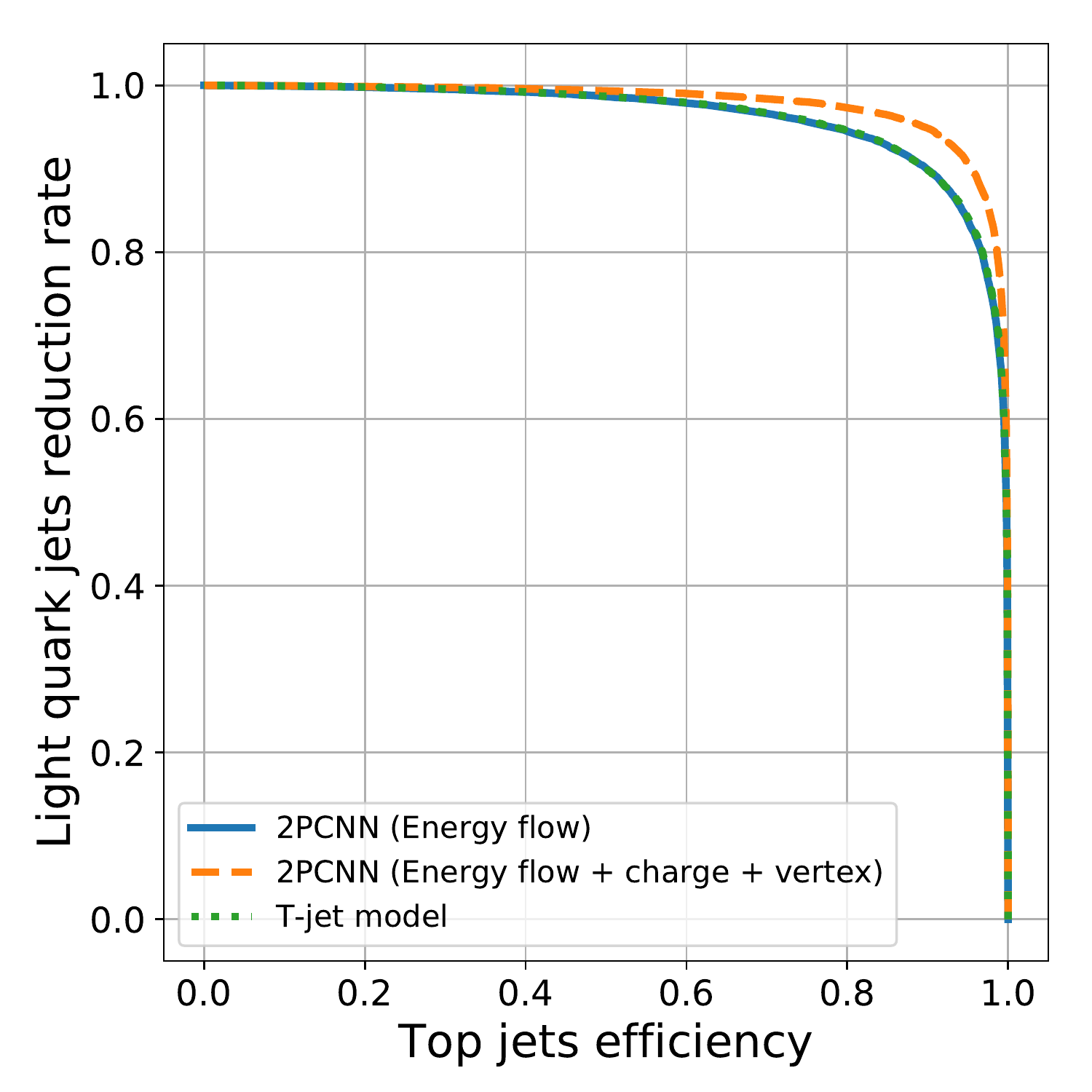}
\caption{The receiver operating characteristic curves for classification of Higgs jets versus light quark jets (left), 
and top jets versus light quark jets (right). The solid curves show the performance of the 2PCNN model based on energy flow information. 
The dashed curves correspond to the 2PCNN model with additional electric charges and vertex inputs. The dotted curves give 
the result from the T-jet model.}
\label{fig:ROC}
\end{center}
\end{figure}

In order to benchmark the 2PCNN performance, we compare with a deep neural network model based on telescoping deconstruction of energy flow information (referred to as the T-jet model) \cite{Chien:2013kca,Chien:2014hla,Chien:2017xrb,Chien:2018dfn}. The method systematically decomposes jet information into a fast-converging subjet series expansion $\sum_N {\rm T}_N$ which is ordered by the number of subjets $N$.  These subjets are defined as the sets of particles along dominant energy flow directions within a variable subjet radius. Such organization is motivated by the infrared structure of QCD. Energetic, collinear particles are captured at lower orders, and the series gradually reaches out to soft, wide-angle particles. In this paper, the T-jet model includes jet information up to the $T_3$ order and scans energy flows with 4 values of subjet radius. The energy flow directions and subjet kinematics consist of 60 input variables. Together with the jet kinematic information, these inputs are processed by a fully connected network layer of 128 nodes followed by two output nodes. The same activation function, loss function and optimizer are adopted as in the 2PCNN model.


FIG.~\ref{fig:ROC} shows the receiver operating characteristic (ROC) curves, plotting background rejection rate as a function of signal efficiency, for two discrimination tasks as representative examples: high $p_T$ Higgs jet versus light quark jet, as well as top jet versus light quark jet. The model performances are quantified by the area under ROC curve (AUC) and the average accuracy (ACC), which is the fraction of correctly-predicted jet samples. 
As summarized in TABLE~\ref{tab:performance}, the 2PCNN and the T-jet model based on energy flow information show nearly the same performance, confirming the baseline capability of the 2PCNN model which is comparable to the state-of-the-art methods that are all capable of modeling the energy flow probability distributions very well. With the additional vertex and charge information, the 2PCNN model achieves excellent performance in all the classification tasks. The vertex information has a strong impact on tagging jets which contain one or more secondary vertices such as the high $p_T$ Higgs and top jets. The electric charges of particles are also essential for separating jets from $W^+$ and $W^-$ bosons.

\begin{table}[t]
\begin{center}
\begin{tabular}{lcccccc}
\hline\hline
& \multicolumn{2}{c}{2PCNN(E-flow)} & \multicolumn{2}{c}{2PCNN(full)} & \multicolumn{2}{c}{T-jet model} \\
Task & ACC & AUC & ACC & AUC & ACC & AUC \\
\hline
$W$ vs quark & 0.881 & 0.945 & 0.881 & 0.946 & 0.880 & 0.945 \\
Higgs vs quark & 0.873 & 0.939 & 0.959 & 0.993 & 0.866 & 0.934 \\
top vs quark & 0.900 & 0.962 & 0.929 & 0.978 & 0.900 & 0.963 \\
$W^+$ vs $W^-$ & 0.505 & 0.502 & 0.757 & 0.839 & 0.502 & 0.502 \\
quark vs gluon & 0.738 & 0.810 & 0.748 & 0.823 & 0.732 & 0.802 \\
\hline\hline
\end{tabular}
\end{center}
\caption{The performance of the 2PCNN and T-jet models, as quantified by the average accuracy (ACC) 
and the area under the receiver operating characteristic curve (AUC), for $W$, Higgs and top tagging as well as $W^+$ versus $W^-$ and quark versus gluon discrimination. The energy flow 2PCNN model has comparable performance with the T-jet model. The 2PCNN model with additional information of electric charges and vertex of charged tracks outperforms significantly the other two models in most of the tasks. The uncertainty due to finite sample size in ACC is smaller than 0.003.}
\label{tab:performance}
\end{table}%

We now discuss the physics properties of the 2PCs and focus on the task of $W$ jet and light quark jet separation using the energy flow 2PCNN model, aiming to identify the key features which are useful for distinguishing the two jet samples. Many other detailed studies will be presented in a forth coming paper. Thanks to the internal ranking of 2PC pairs, the importance of the top-$k$ ranked 2PC pairs within a filter can potentially be quantified by their filter output values. These sets of outputs represent the weights on 2PCs which the 2PCNN has learned from separating the two samples and are task-dependent. Therefore intrinsic features of each jet sample can be illuminated by contrasting with different jet samples potentially having distinct features. 

\begin{figure}[t]
\begin{center}
\includegraphics[width=0.23\textwidth]{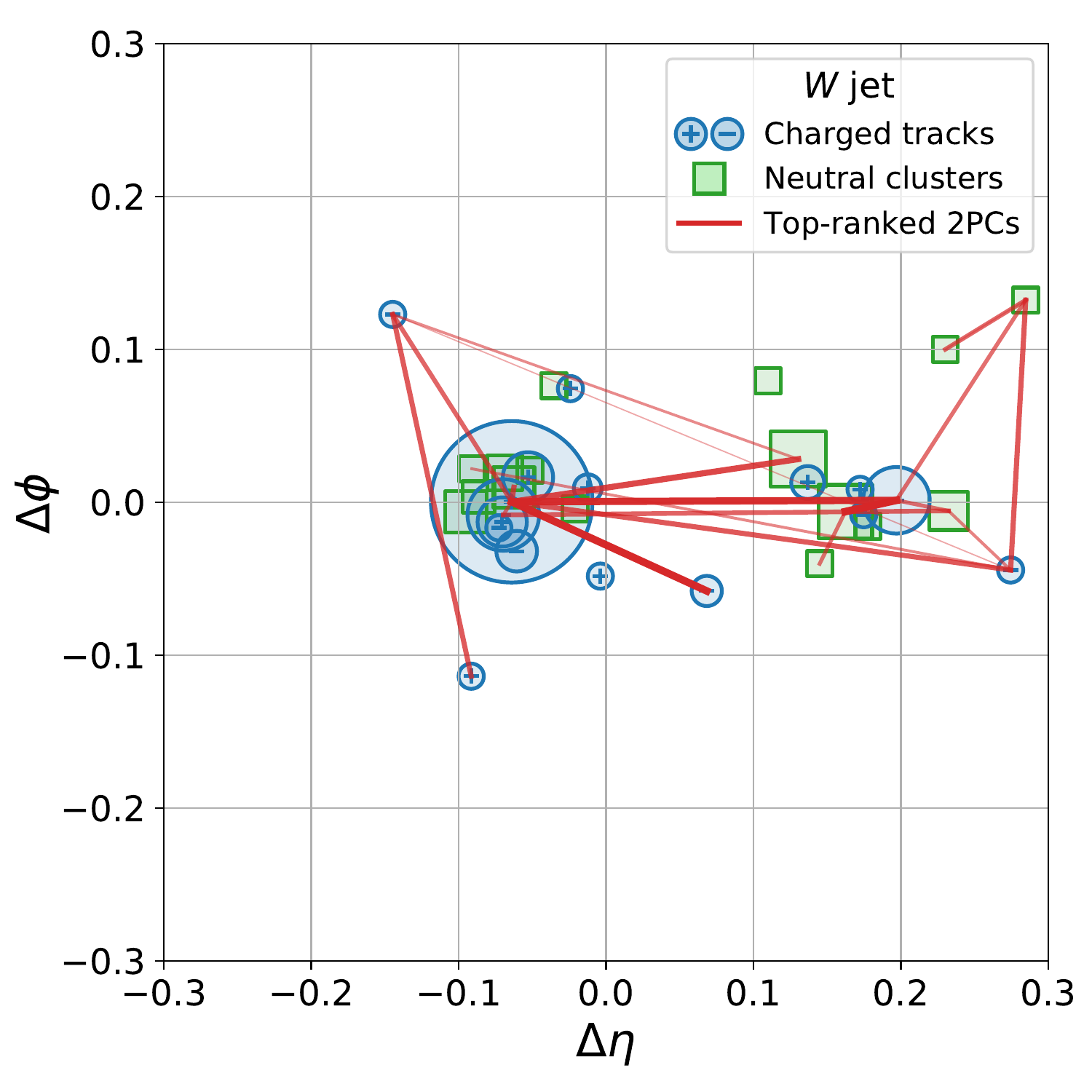}
\includegraphics[width=0.23\textwidth]{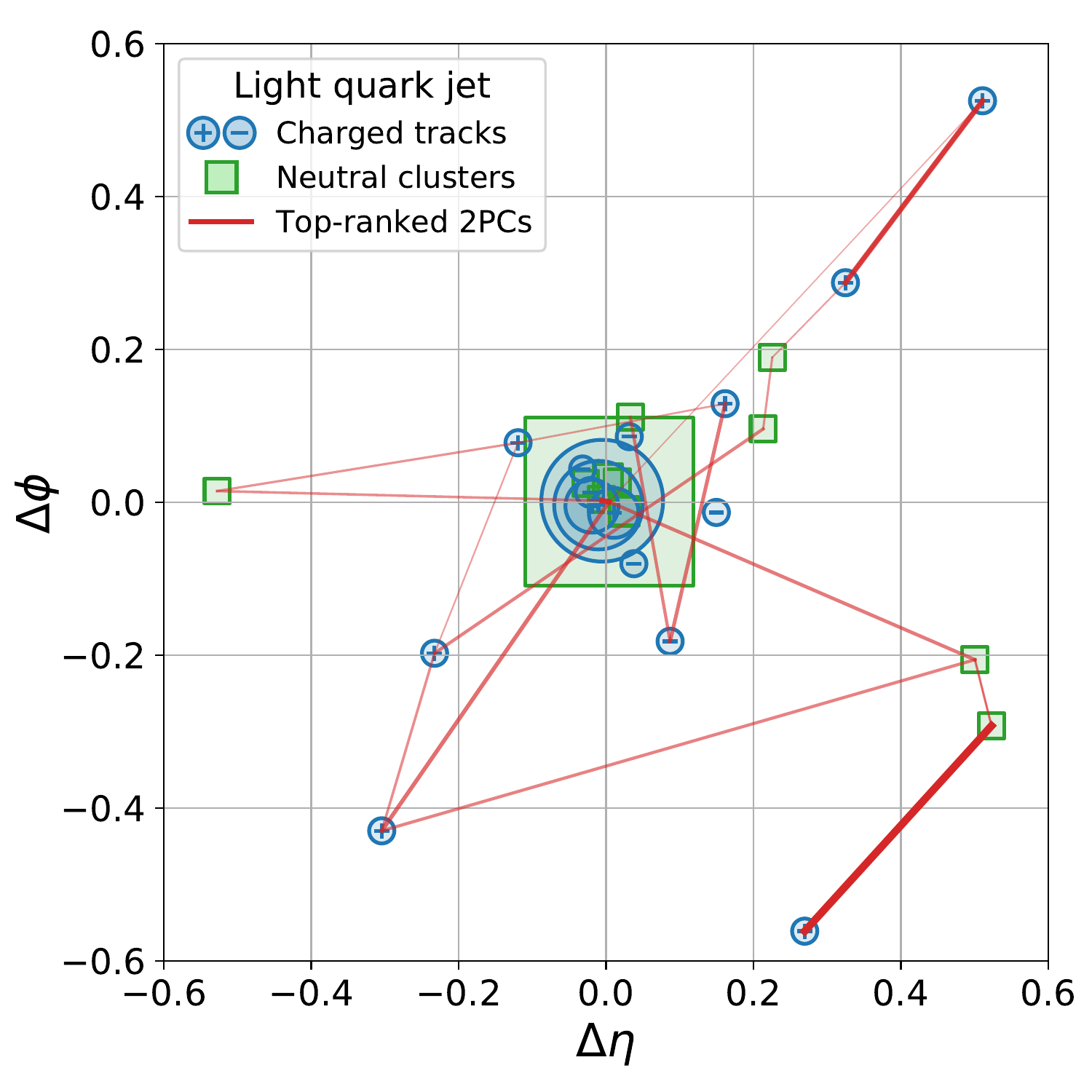}
\caption{Displays of a typical $W$ jet (left) and a typical light quark jet (right) in $\Delta\eta$-$\Delta\phi$ plane. 
The charged tracks of jet particles are shown as circles with charge signs, while the neutral clusters are shown as squares. 
The sizes of the circles or the squares are proportional to the $p_T$'s of jet constituents. 
The solid lines indicate the top-one ranked 2PCs of the filters in the energy flow 2PCNN model. 
The strength of filter outputs are represented by the line thickness.}
\label{fig:jet_disp}
\end{center}
\end{figure}

FIG.~\ref{fig:jet_disp} shows the display of a typical two-prong $W$ jet and a typical one-prong light quark jet. The jet constituents are shown as scattered circles and squares, with their sizes proportional to the particle transverse momenta. The top-one ranked 2PC pair of each active 2PCNN filter is indicated by a solid line, with the thickness representing the strength of the filter output. Two distinct signatures of the high-ranked 2PCs are identified:
(1) strong internal correlations within and between the prongs, and 
(2) strong correlations between high $p_T$ constituents within the prongs and low $p_T$ constituents scattered at wide angle. 


\begin{figure*}[t]
\begin{center}
\includegraphics[width=0.246\textwidth]{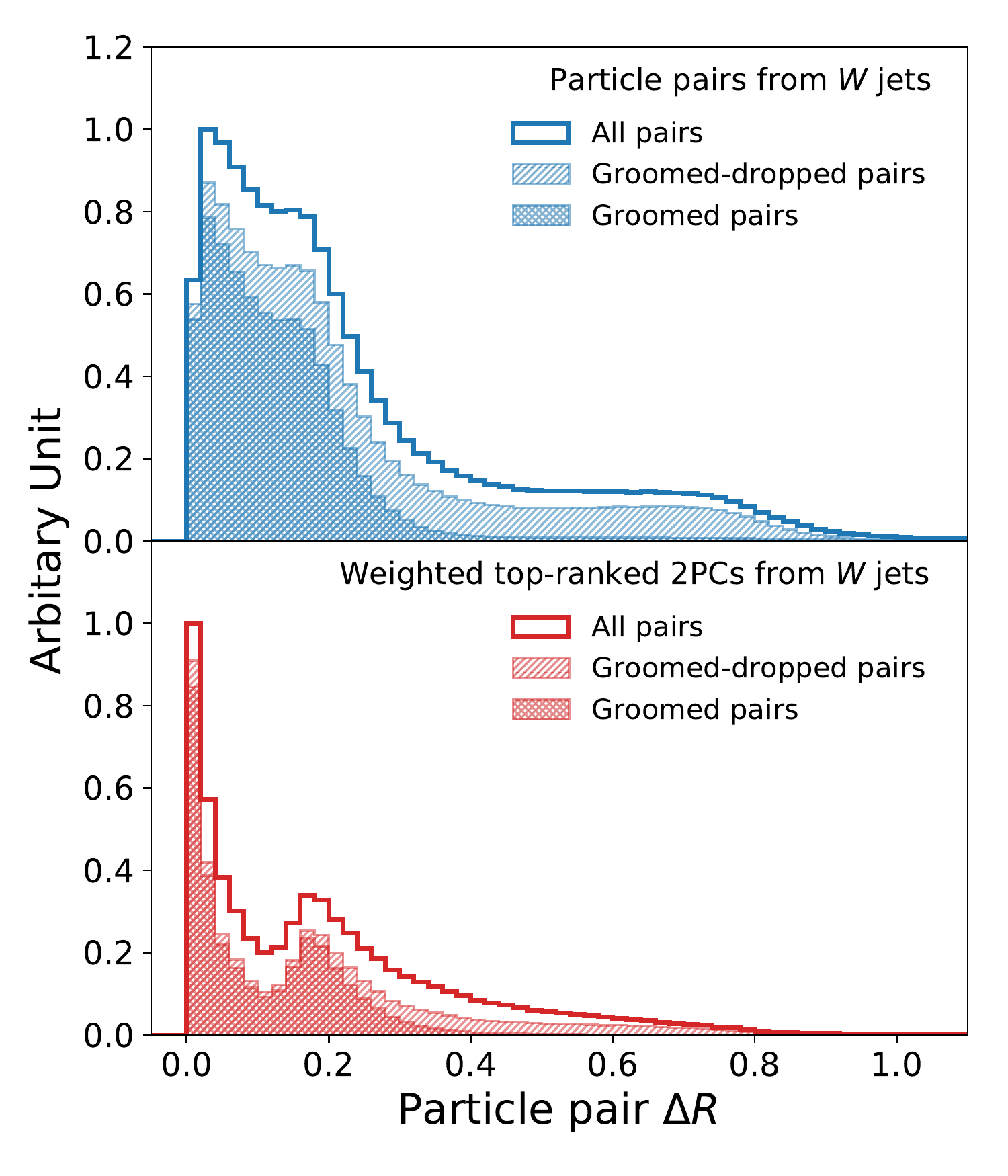}
\includegraphics[width=0.246\textwidth]{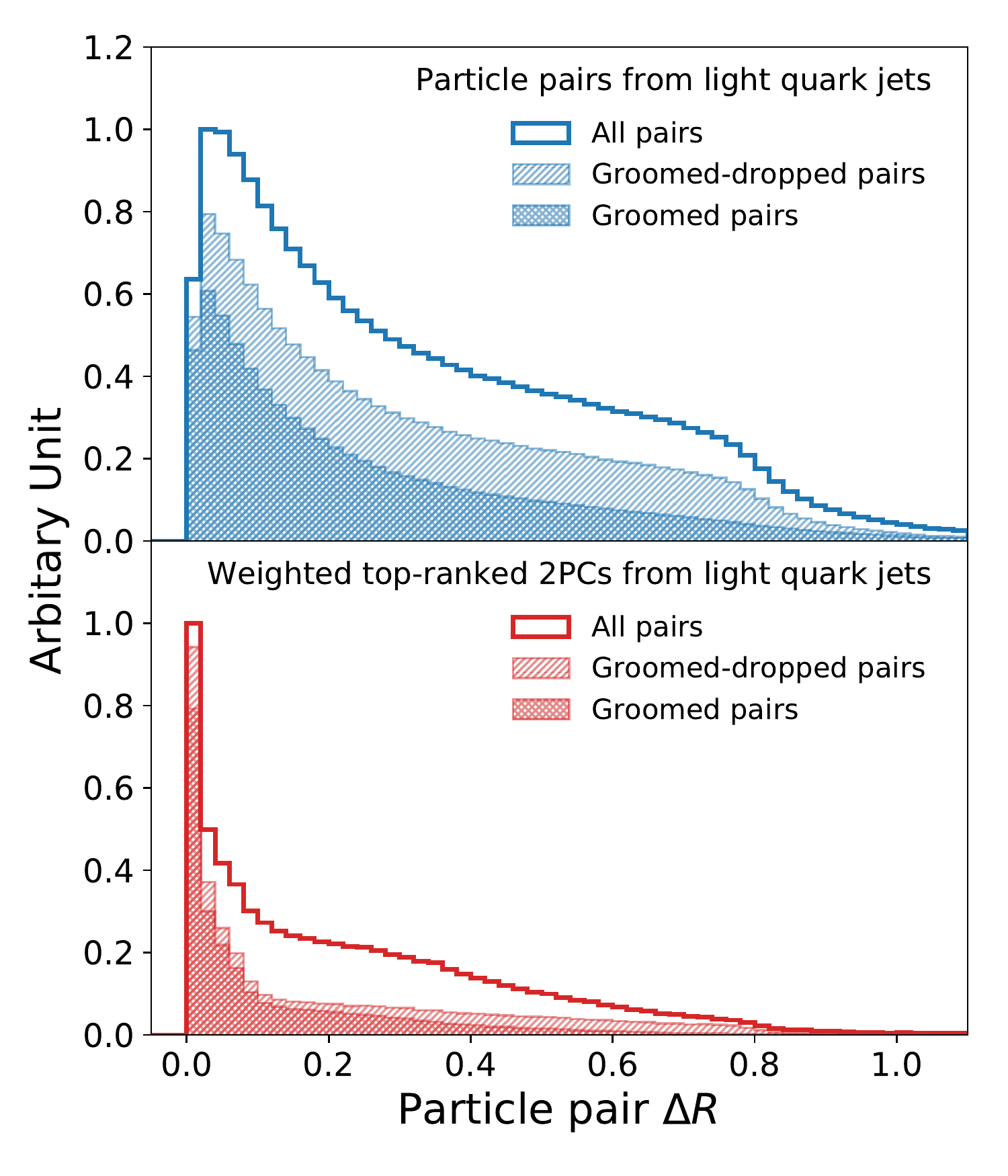}
\includegraphics[width=0.246\textwidth]{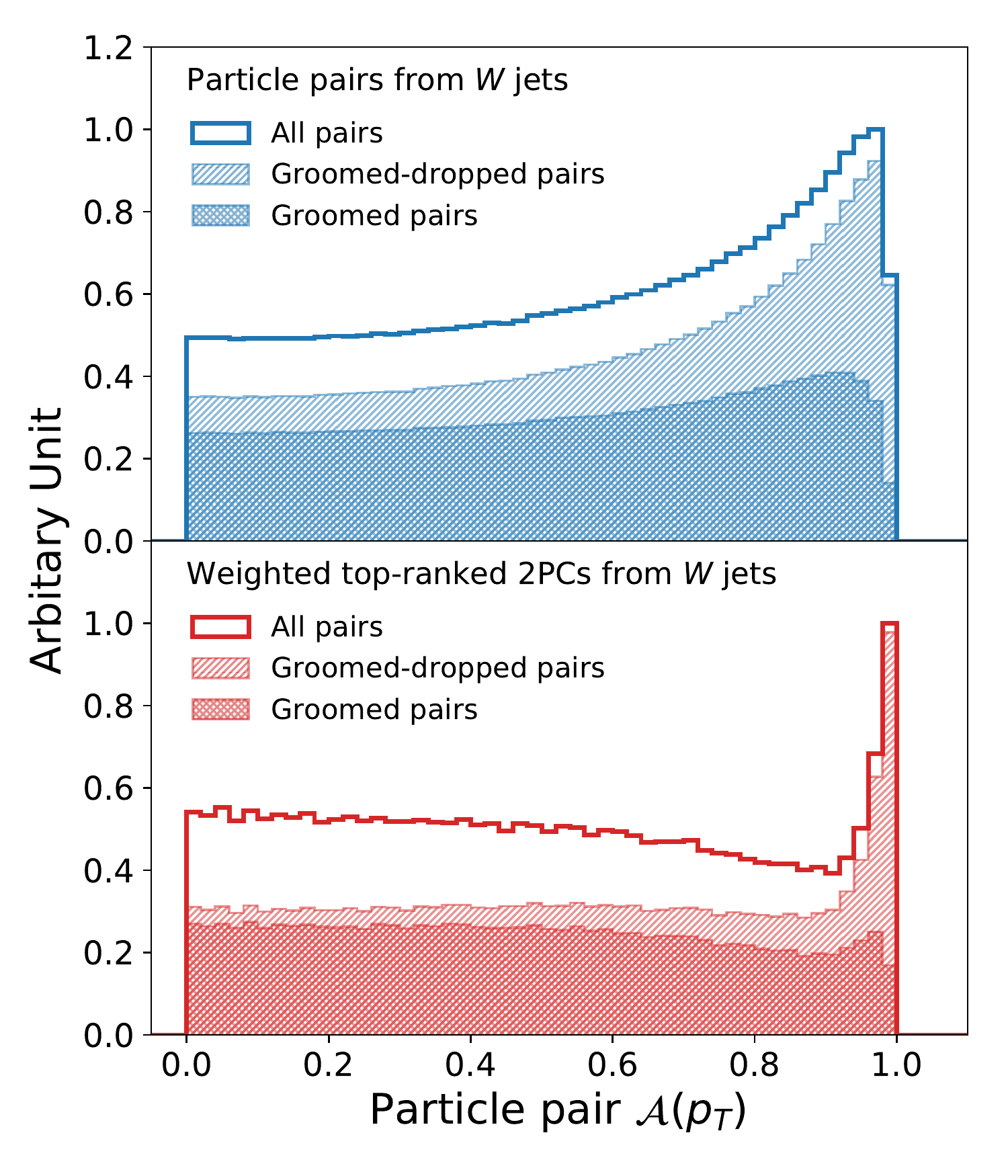}
\includegraphics[width=0.246\textwidth]{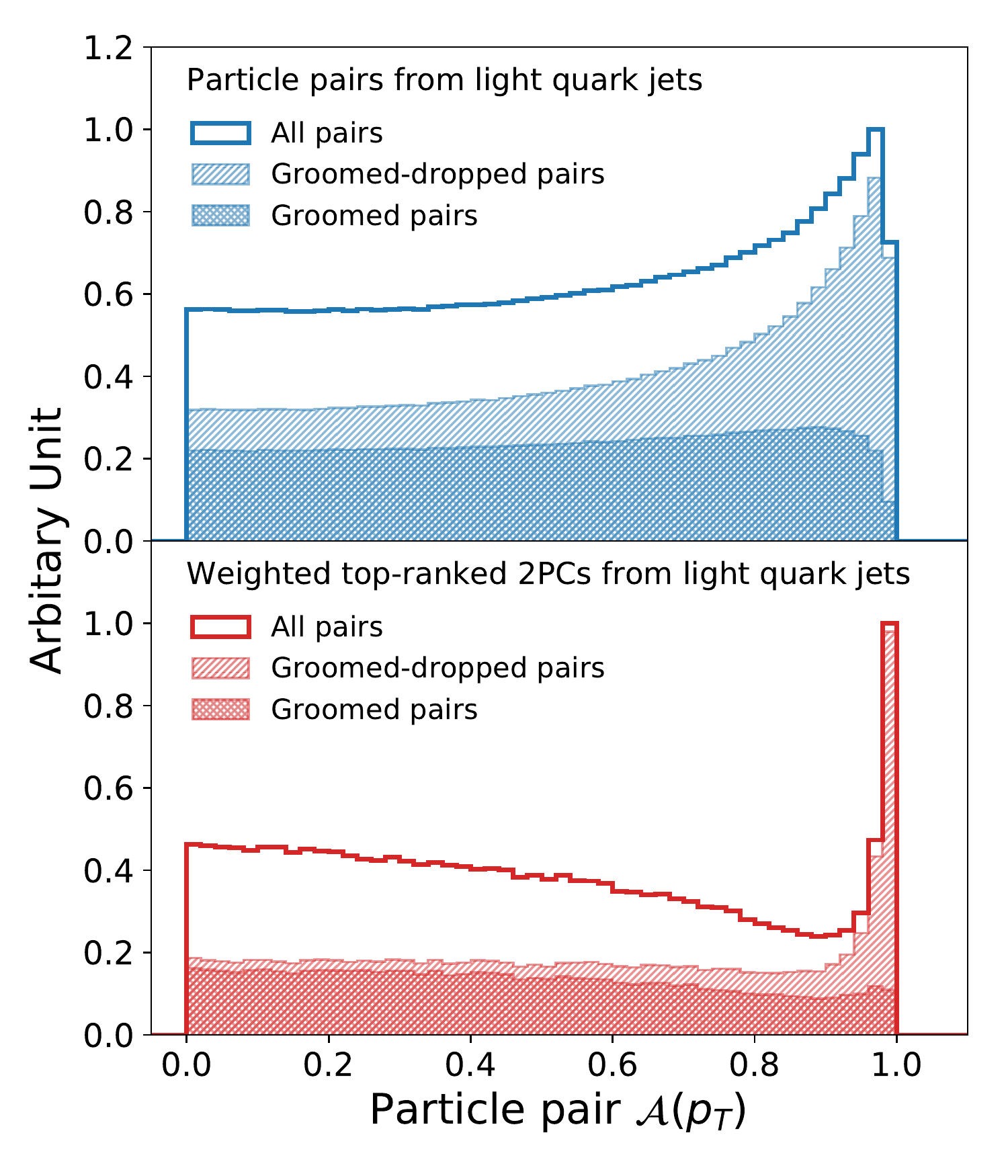}
\caption{The distributions of spatial distance $\Delta R$ (left two panels) and $p_T$ asymmetry $\mathcal{A}(p_T)$ (right two panels) between particles in 2PCs for $W$ jets and light quark jets. The lower panels show the distributions for top-ranked 2PCs weighed by the 2PCNN filter outputs, while the upper panels have equal weights for all the 2PCs. In each subfigure, the solid lines represent the histograms from all 2PC pairs, which are decomposed into three stacked, hatched or unfilled components corresponding to groomed-groomed, groomed-dropped and dropped-dropped 2PC pairs as categorized by soft-drop and collinear-drop methods. }
\label{fig:dr_ptasy}
\end{center}
\end{figure*}
 
Such behaviors of the high-ranked 2PC pairs are further examined by the spatial distance $\Delta R = \sqrt{(\eta^i-\eta^j)^2+(\phi^i-\phi^j)^2}$ between the $i$-th and $j$-th particles forming the 2PC, and their $p_T$ asymmetry $\mathcal{A}(p_T) = |p_T^i-p_T^j|/(p_T^i+p_T^j)$. FIG.~\ref{fig:dr_ptasy} shows the comparisons of a variety of $\Delta R$ and $\mathcal{A}(p_T)$ distributions of $W$ jets and light quark jets. In order to maximize the sensitivity to the features extracted by the 2PCNN, the distributions corresponding to the top-ranked 2PCs weighed by the output values of 2PCNN filters, as an indication of their importance, are presented in the lower panels. For $W$ jets, strong features are identified at $\Delta R\approx 0$ and $\Delta R\approx 0.2\sim 2m_W/p_T({\rm jet})$, whereas for light quark jets the $\Delta R\approx 0$ feature is strong and the $\Delta R\approx 0.2$ feature is absent. This indicates the intrinsic jet property of particle collimation for both samples, and the two-prong structure of $W$ jets. The filters tend to either select the 2PCs within the same prong therefore with small $\Delta R$ values, or emphasize the correlations between the two prongs for $W$ jets and build up the $\Delta R\approx 0.2$ feature. On the other hand, a clear feature at $\mathcal{A}(p_T)\approx 1$ shows up in the filter-output weighed $\mathcal{A}(p_T)$ distributions for both samples. Such signature corresponds to highly unbalanced $p_T$'s in the 2PCs therefore one of the particle has to be soft. This shows the importance of low $p_T$ constituents which are often neglected or suppressed in many other jet tagging methods.

In order to further examine the properties of 2PCs which are responsible for the learned jet features, soft-drop and collinear-drop with parameters $z_{\rm cut}=0.2$ and $\beta=0$ are used to classify jet constituents into two categories. The jet constituents surviving soft-drop are referred to as ``groomed," while those surviving collinear-drop belong to the ``dropped" category. Therefore the 2PCs form three distinct sets: groomed-groomed, groomed-dropped and dropped-dropped. We can see that the one and two-prong structures are dominantly determined by the groomed-groomed 2PC pairs from the two soft-drop branches. Also, there is a significant dropped-dropped contribution at medium and large $\Delta R$ values for light quark jets. On the other hand, the feature at $\mathcal{A}(p_T)\approx 1$ dominantly comes from the groomed-dropped 2PC pairs which correlate hard, collinear particles to soft, wide-angle particles, while most other 2PC pairs form a fairly flat $\mathcal{A}(p_T)$ distribution. 

To highlight the power and sensitivity of 2PCNN in feature extraction, we contrast with the $\Delta R$ and $\mathcal{A}(p_T)$ distributions of $W$ and light-quark jets formed with equal weight for all the 2PC pairs (upper panels of FIG.~\ref{fig:dr_ptasy}). Evidence of one- or two-prong structure from the falling $\Delta R$ distribution with a ``shoulder" around $\Delta R \approx 0.2$, as well as the significant soft particle contributions in the $\mathcal{A}(p_T)\approx 1$ region, are observed. Similar conclusions can be reached by decomposing the distributions into groomed-groomed, groomed-dropped and dropped-dropped components; however all the features are much more convincingly identified by the 2PCNN model as a very useful guide for physics analysis. 


In conclusion, we have constructed a new neural network architecture which utilizes two-particle correlations (2PCs) as a fundamental description of jets. The input for 2PCNN is dynamically determined by the number of jet constituents with no artificial reduction of input information and no particular biased ordering of jet particles. The structure of the 2PC neural network is driven by the physics needs, rather than a direct application of existing deep learning methods developed for solving problems in other subjects. We demonstrate that the 2PCNN model based on energy flow information has comparable performance with the model using variables from telescoping deconstruction, which is one of the most effective method for factorizing jet information. 
By including additional information from charged tracks, such as electric charges and vertex, the 2PCNN model achieves an unprecedentedly promising power for a variety of jet tagging tasks. Besides the excellent tagging performance, an important benefit of the 2PCNN model is the ranking of 2PCs which can be directly extracted from the filter outputs. Since two-particle correlations are fundamental descriptions of particles relations, this physical machine-learning method can be potentially useful in subsequent physics studies such as hadronization process and collective behaviors of quark-gluon plasma remnants in high energy collisions. The 2PCNN will shed light on physics signatures which are difficult to identify with conventional methods.

\section{Acknowledgements}
The authors thank George Sterman for encouraging and useful conversations, as well as Cheng-Wei Chiang, Fr\'ed\'eric Dreyer, Sung Hak Lim, Matthew Schwartz and Jesse Thaler for helpful comments and suggestions. Y.-T. Chien was supported by the National Science Foundation grant PHY-1915093. 
K.-F. Chen was supported by the grant 106-2112-M-002-006 of Ministry of Science and Technology, Taiwan. 

\bibliography{ref}

\end{document}